# Hanbury Brown Twiss effect for ultracold quantum gases

M. Schellekens,[1] R. Hoppeler,[1] A. Perrin,[1] J. Viana Gomes,[1,2]
D. Boiron,[1] A. Aspect,[1] C. I. Westbrook[1*]

[1]Laboratoire Charles Fabry de l'Institut d'Optique, UMR 8501 du CNRS
Centre Scientifique d'Orsay, Bat. 503, 91403 Orsay CEDEX France
[2]Departimento de Fisica, Universidade do Minho
4710-057, Braga Portugal

[*]To whom correspondence should be addressed. E-mail: christoph.westbrook@iota.u-psud.fr.

**We have studied 2-body correlations of atoms in an expanding cloud above and below the Bose-Einstein condensation threshold. The observed correlation function for a thermal cloud shows a bunching behavior, while the correlation is flat for a coherent sample. These quantum correlations are the atomic analogue of the Hanbury Brown Twiss effect. We observe the effect in three dimensions and study its dependence on cloud size.**

Nearly half a century ago, Hanbury Brown and Twiss (HBT) performed a landmark experiment on light from a gaseous discharge (*1*). The experiment demonstrated strong correlations in the intensity fluctuations at two nearby points in space despite the random or chaotic nature of the source. Although the effect was easily understood in the context of classical statistical wave optics, the result was surprising when viewed in terms of the quantum theory. It implied that photons coming from widely separated points in a source such as a star were "bunched". On the other hand, photons in a laser were not bunched (*2,3*). The quest to understand the observations stimulated the birth of modern quantum optics (*4*). The HBT effect has since found applications in many other fields from particle physics (*5*) to fluid dynamics (*6*).

Atom or photon bunching can be understood as a two-particle interference effect (*7*). Experimentally, one measures the joint probability for two particles, emitted from two separated source points $A$ and $B$, to be detected at two detection points, $C$ and $D$. One must consider the quantum mechanical amplitude for the process ($A \to C$ and $B \to D$) as well as that for ($A \to D$ and $B \to C$). If the two processes are indistinguishable, the amplitudes interfere. For bosons, the interference is constructive resulting in a joint detection probability which is enhanced compared to that of two statistically independent detection events, while for fermions the joint probability is lowered. As the detector separation is increased, the phase difference between the two amplitudes grows large enough that an average over all possible source points $A$, $B$ washes out the interference and one recovers the situation for uncorrelated detection events. This fact was used by HBT to measure the angular size of a star (*8*), but another major consequence of the observation was to draw attention to the importance of two-photon amplitudes, and how their interference can lead to surprising effects. These quantum amplitudes must not be confused with classical electromagnetic field amplitudes (*3*). Two-photon states subsequently led to many other striking examples of "quantum weirdness" (*9*). In contrast to a chaotic source, all photons in a single mode laser are in the same quantum state. Hence there is only one phys-ical process and no bunching effect. A similar effect is expected for atoms in a Bose-Einstein Condensate (BEC).

Two-particle correlations have been observed both for cold neutral atoms (*10–12*) and for electrons (*13–15*), and three-particle correlations (*16–18*) at zero distance have also been used to study atomic gases. But the full three dimensional effect and its dependence on the size and degeneracy of a sample has yet to be demonstrated for massive particles. Here we demonstrate the effect for a trapped cloud of atoms close to the BEC transition temperature released onto a detector capable of individual particle detection. We extract, for varying cloud sizes, a three-dimensional picture of the correlations between identical particles produced by quantum interference. We also show that a BEC shows no such correlations. The results are in agreement with an ideal gas model and show the power of single particle detection techniques applied to the study of degenerate quantum gases.

The calculation of the phase difference of the possible two-particle detection amplitudes given in (*7*) can be adapted to the case of particles of mass $m$ travelling to a detector in a time $t$. One can show that the correlation length observed at the detectors, *i.e.* the typical detector separation for which interference survives, is $l_i = \dfrac{\hbar t}{m s_i}$ where $s_i$ is the source size along the direction $i$, $\hbar$ is the reduced Planck's constant and we have assumed that the size of the cloud at the detector is much larger than the initial size. The optical analog of this expression, for a source of size $s$ and wavelength $\lambda$ at a distance $L$ from the observation plane, is $l = \dfrac{L\lambda}{2\pi s}$. This is the length scale of the associated speckle pattern. The formula can be recovered for the case of atoms travelling at constant velocity $v$ towards a detector at distance $L$ if one identifies $h/mv$ with the deBroglie wavelength corresponding to velocity $v$. The formula we give is also valid for atoms accelerated by gravity. The interpretation of $l$ as the atomic speckle size remains valid. A pioneering experiment on atom correlations used a continuous beam of atoms (*10*). For a continuous beam, the correlation time, or equivalently, the longitudinal correlation length, depends on the velocity width of the source and not on the source size. Thus, the longitudinal and transverse directions are qualitatively different. By contrast, our measurements are performed on a cloud of atoms released suddenly from a magnetic trap. In this case, the 3 dimensions can all be treated equivalently and the relation above applies in all three. As the trap is anisotropic, the correlation function is as well, with an inverted ellipticity. Our sample is a magnetically trapped cloud of metastable helium atoms evaporatively cooled close to the BEC transition temperature (*19*) (about 0.5 $\mu$K for our conditions). Our source is thus very small and together with a long time of flight (308 ms) and helium's small mass, we achieve a large speckle size or correlation volume ($30 \times 800 \times 800$ $\mu$m$^3$) which simplifies the detection problem. For example the observations are much less sensitive to the tilt of the detector than in (*10*).

To detect the atoms we use an 8 cm diameter microchannel plate detector (MCP). It is placed 47 cm below the center of the magnetic trap. A delay line anode permits position sensitive detection of individual particles in the plane of the detector (*20*) (Fig. 1). Atoms are released from the trap by suddenly turning off the magnetic field. Approximately $10\%$ of these atoms are transferred to the magnetic field insensitive $m = 0$ state by non-adiabatic transitions (*19*) and fall freely to the detector. The remaining atoms are removed by applying additional magnetic field gradients during the time of flight. For each detected atom we record the in-plane coordinates $x$, $y$ and the time of detection $t$. The atoms hit the detector at 3 m/s with a velocity spread below $1\%$ and so we convert $t$ into a vertical position $z$. The observed rms resolution is $d \sim 250$ $\mu$m in $x$ and $y$ and 2 nm in $z$. These data allow us to construct a 3 dimensional histogram of pair separations ($\Delta x$, $\Delta y$, $\Delta z$) for all particles detected in a single cloud. The histograms are summed over the entire atomic distribution and over many shots, typically 1000 (*21*).

Because of our good resolution along $z$, we begin by concentrating on the correlation function along this axis. Normalized correlation functions for various experimental conditions are shown in Fig. 2A. To compute the normalized correlation function, we divide the pair separa-tion histogram by the auto-convolution of the average single particle distribution along $z$. We also normalize the correlation function to unity for large separations. This amounts to dividing, for each elementary pixel of our detector, the joint detection probability by the product of the individual detection probabilities at the two pixels. This gives us the usual normalized correlation function $g^{(2)}(\Delta x = 0, \Delta y = 0, \Delta z)$. The HBT bunching effect corresponds to the bump in the top 3 graphs of Fig. 2A. The fourth graph shows the result for a BEC. No correlation is observed. (A detector saturation effect in the BEC data required a modified analysis procedure (*21*).) We have also recorded data for a cloud with a 2 mm radius and 1 mK temperature, for which the correlation length is so small that the bunching effect is washed out by the in-plane detector resolution. Experimentally, the normalized correlation function in this case is indeed flat to within less than 1 %.

We plot in Fig. 2B the normalized correlation functions in the $\Delta x - \Delta y$ plane and for $\Delta z = 0$, for the same three data sets. The data in Fig. 2B clearly show the asymmetry in the correlation function arising from the difference in the two transverse dimensions of the trapped cloud. The long axis of the correlation function is orthogonal to that of the magnetic trap.

We expect the experimental normalized correlation function for a thermal bosonic gas to be described by:

$$g^{(2)}_{\text{th}}(\Delta x, \Delta y, \Delta z) = 1 + \eta \exp\left(-\left[\left(\frac{\Delta x}{l_x}\right)^2 + \left(\frac{\Delta y}{l_y}\right)^2 + \left(\frac{\Delta z}{l_z}\right)^2\right]\right) \quad (1)$$

We have assumed here that the gas is non interacting and that the velocity and density distribution remain roughly Gaussian even close to the BEC transition temperature. Numerical simulations indicate that this is a good approximation when the correlation function is averaged over the entire cloud (*22*). As discussed above, the correlation lengths should be inversely proportional to the sizes $s_i$ of the sample. In a harmonic trap with trapping frequency $\omega_i$ along the $i$ direction one has $s_i = \sqrt{\frac{k_B T}{m \omega_i^2}}$ where $k_B$ is Boltzmann's constant and $T$ is the temperature of the atoms. Since $T$ is derived directly from the time of flight spectrum, we shall plot our data as a function of $T$ rather than of $s$. The parameter $\eta$ would be unity for a detector whose resolution width $d$ is small compared to the correlation length. Our $d$ is smaller than $l_y$ but larger than $l_x$ and in this case the convolution by the detector resolution results in an $\eta$ given approximately by $\frac{l_x}{2d} \sim 5\%$. We use Eq. 1 to fit the data using $\eta$ and the $l_i$ as fit parameters, and compare the results to the ideal gas model (*21*).

The results for $l_x$, $l_y$ and $l_z$ for our three temperatures are plotted in Fig. 3A. The fitted values of $l_x$ are $\sim 450$ $\mu$m, and are determined by the detector resolution rather than the true coherence length along $x$. The value of $l_y$ has been corrected for the finite spatial resolution of the detector. The fitted value of $l_z$ requires no correction since in the vertical direction the resolution of the detector is much better. One sees that $l_y$ and $l_z$ are consistent and agree with the prediction using the known trap frequencies and temperatures. Fig. 3B shows the fitted value of $\eta$ versus temperature, along with the prediction of the same ideal gas model as in Fig. 3A and using the measured detector resolution. The data are in reasonable agreement with the model although we may be seeing too little contrast at the lowest temperature. The run at 0.55 $\mu$K was above but very close to the BEC transition temperature. (We know this because when taking data at 0.55 $\mu$K about one third of the shots contained small BECs; these runs were eliminated before plotting Fig. 2.) Future work will include examining whether the effect of the repulsive interactions between atoms or finite atom number must be taken into account.

The results reported here show the power of single particle detection in the study of quantum gases. The correlations we have observed are among the simplest which should be present. Two recent experiments have shown correlations in a Mott insulator (*11*) as well as in atoms produced from the breakup of molecules near a Feshbach resonance (*12*). Improved observations of these effects may be possible with individual particle detection. Other atom pair production mechanisms, such as 4-wave mixing (*23, 24*) can be investigated. A fermionic analog to this experiment using $^3$He would also be (*25*) of great interest.

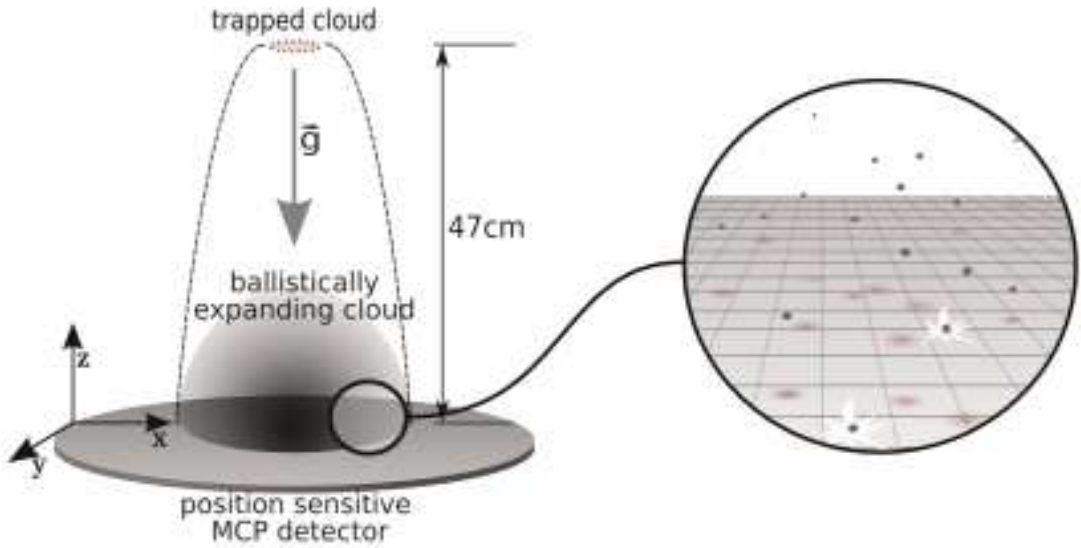

**Fig. 1.** Schematic of the apparatus. The trapped cloud has a cylindrical symmetry with oscillation frequencies $\omega_x/2\pi = 47$ Hz and $\omega_y/2\pi = \omega_z/2\pi = 1150$ Hz. During its free fall towards the detector, a thermal cloud acquires a spherical shape. A 1 $\mu$K temperature yields a cloud with an rms radius of approximately 3 cm at the detector. Single particle detection of the neutral atoms is possible because of the atom's 20 eV internal energy that is released at contact with the MCP. Position sensitivity is obtained through a delay-line anode at the rear side of the MCP.

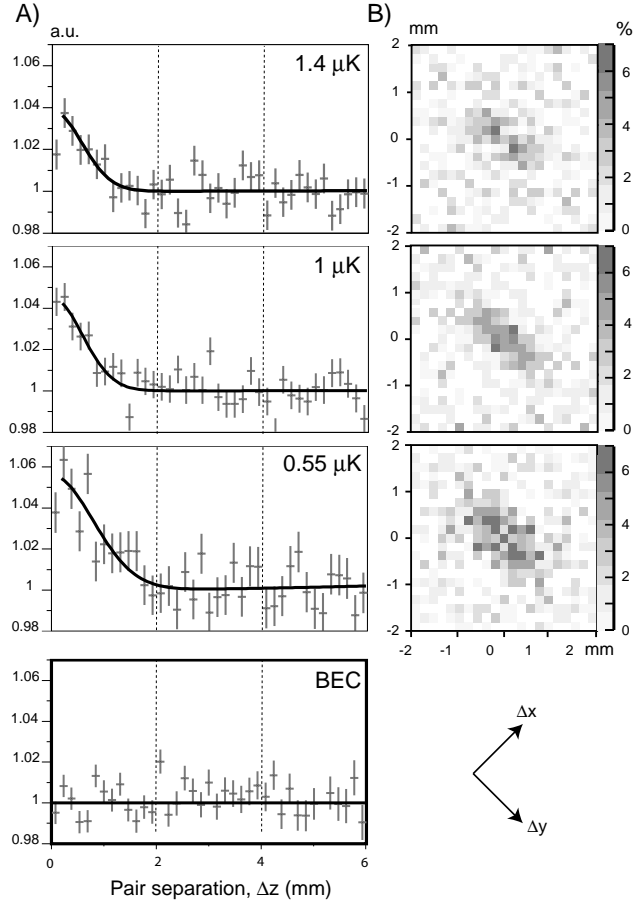

**Fig. 2.** A) Normalized correlation functions along the vertical ($z$) axis for thermal gases at 3 different temperatures and for a BEC. For the thermal clouds, each plot corresponds to the average of a large number of clouds at the same temperature. Error bars correspond to the square root of the number of pairs. B) Normalized correlation functions in the $\Delta x - \Delta y$ plane for the 3 thermal gas runs. The arrows at the lower right show the $45°$ rotation of our coordinate system with respect to the axes of the detector. The inverted ellipticity of the correlation function relative to the trapped cloud is clearly visible.

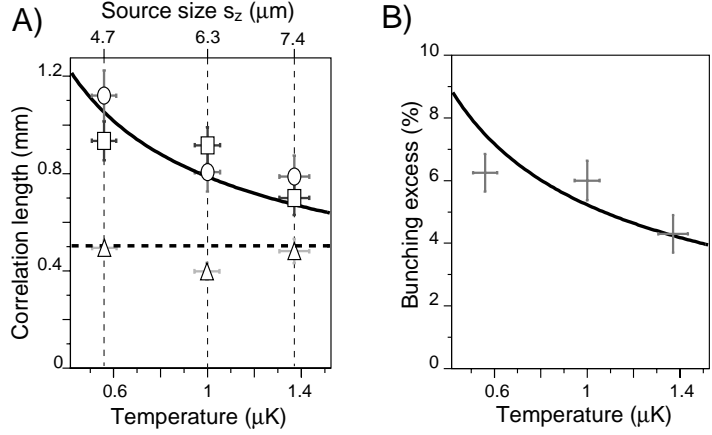

**Fig. 3.** Results of fits to the data in Figs 2A, 2B. A) Fitted correlation lengths $l_x, l_y$ and $l_z$ along the 3 axes (triangles, squares and circles) as a function of temperature. The upper axis shows the corresponding source size $s_z$. Vertical error bars are from the fits. Horizontal error bars correspond to the standard deviation of the measured temperature. Along the $x$ axis the measurement is entirely limited by the detector resolution. The dotted horizontal line is the result of an independent estimate of the resolution. The result for the $y$ axis has been corrected for the finite detector resolution as characterized by the fitted value of $l_x$. The $z$-axis suffers from no such resolution limit. The solid curve corresponds to $\dfrac{\hbar t}{m s_z}$. B) Fitted contrast $\eta$ of the correlation function for the three temperatures used. The solid line corresponds to the same non-interacting gas model as the line in A) (*21*) and includes the finite detector resolution.

# Supplementary online material

**Histogram along** $z$ : In computing the vertical separation histogram (the unnormalized correlation function), our averaging procedure is as follows. Let the index $(i,j)$ denote a particular detector pixel in the $x - y$ plane. We compute the histogram $h_{i,j}$, corresponding to vertical separations of the pairs of atoms detected in the pixel $(i,j)$ as well as the histogram $h_{i,j;k,l}$, corresponding to the separations of pairs of atoms one of which was in pixel $(i,j)$ and the other in pixel $(k,l)$. To relate pixel index to position we use the 200 $\mu$m pixel size (not the same as the resolution $d$) and the fact that the axes of the detector are at $45°$ to the trap axes. This gives $\frac{1}{\sqrt{2}}(\Delta x + \Delta y) = (k - i) \times 200$ $\mu$m and $\frac{1}{\sqrt{2}}(\Delta x - \Delta y) = (l - j) \times 200$ $\mu$m. To improve the signal to noise ratio in the correlation function along the z axis we form the sum:

$$h_{i,j} + \sum_{a,b} h_{i,j;i+a,j+b} \qquad (2)$$

in which $(a,b) = \{(0,-1), (1,0), (1,-1), (1,-2), (2,-1), (2,-2), (2,-3), (3,-2), (3,-3)\}$. This procedure allows us to include more pairs along the y axis for which the correlation length is longer. The histogram is symmetric under inversion by construction. Our choice of pixels for the sum avoids double counting. The resulting histogram is then summed over all pixels $(i,j)$ and all the cloud realizations and is plotted in Fig. 2A. After averaging, a 150 $\mu$m vertical pixel contains typically $2 \times 10^4$ pairs. This averaging procedure is not used in Fig. 2B (see below).

The analysis of the BEC presented an additional complication because the high density of the sample, even after expansion, appeared to induce saturation effects in the detector. When observing a BEC with about 1000 detected atoms, the second half of the cloud was detected with much reduced efficiency. This effect was of course more pronounced at the center of the BEC, leading to "banana' shaped rather than circular profiles in the $y - z$ plane. The saturation effect caused a high sensitivity of the apparent shape of a cloud to the number of atoms in it.

Since our averaging and normalization procedure assume that all shots have the same spatio-temporal shape, the cloud shapes must be corrected before averaging (*i.e.* the "bananas" must be straightened out - or at least all be given the same curvature). This correction was done, on a shot by shot basis, by finding the location of the maximum of the arrival time distribution (recall that the $z$ axis is the temporal axis) as a function of $y$. These maxima were fit to a polynomial and this polynomial was then subtracted from to the $z$ coordinate of all detection events thus yielding an approximately ellipsoidal shape. Atom pairs which are close together remain close under such a transformation despite the distortion produced in the overall shape in the cloud. Thus the correlations, if present, should be substantially preserved.

**Fitting procedure :** Instead of fitting the data directly to the function $g_{\text{th}}^{(2)}$, we use a three step procedure that exploits the Gaussian nature of $g_{\text{th}}^{(2)}$. First, since our resolution is best along the $z$ direction, we fit the data of Fig. 2A to $g_{\text{th}}^{(2)}(0,0,\Delta_z)$ to extract $l_z$. The assumption of a Gaussian $g_{\text{th}}^{(2)}$ ensures that the averaging over pixels as described in Eq. 2 improves the uncertainty in the fitted value of $l_z$ only at the cost of a lowered value of $\eta$. Then, we fit the normalized experimental correlation function for a given value of $(\Delta x, \Delta y)$ without the average over pixels in Eq. 2 to $1 + \eta' \exp[-(\Delta z/l_z)^2]$ and fixing the value of the parameter $l_z$ to that extracted in the previous step. This gives the value of $\eta' = \eta \exp(-[(\Delta x/l_x)^2 + (\Delta y/l_y)^2])$ as a function of $(\Delta x, \Delta y)$ which we plot in Fig. 2B. We then fit the data in Fig. 2B to $g_{\text{th}}^{(2)}(\Delta x, \Delta y, 0)$ to find $l_x$ and $l_y$ and $\eta$.

**Effect of the detector resolution on the correlation function :** In the $x-y$ plane, the detector has a Gaussian resolution of rms width $d$. Thus a two-particle detection probability has a rms resolution of $\sqrt{2}\, d = 350$ $\mu$m. Taking this into account, the normalized two-body correlation function has a contrast of

$$\eta = 1/\sqrt{(1+4d^2/l_x^2)(1+4d^2/l_y^2)} \sim l_x/(2d)$$

This function is plotted as a solid line in Fig. 3B. The finite detector resolution also means that the fitted value of $l_x$ should be $\sqrt{2} \times \sqrt{2}\, d = 2d = 500$ $\mu$m ($l_x$ is not defined as a rms width).